\documentstyle[pra,aps,eqsecnum]{revtex}

\def\ut#1{\lower1.2ex\hbox{$\mathchar"3218$}\mkern -14mu%
          \hbox to 2ex{\hss$#1$\hss}}

\begin{document}
\draft

\title{Collective excitations of degenerate Fermi gases in anisotropic
       parabolic traps}
\author{Andr\'as Csord\'as}
\address{Research Group for Statistical Physics of the Hungarian    
         Academy of Sciences\\
         1117 P\'azm\'any P\'eter S\'et\'any 1/A\\ 
         Budapest, Hungary}
\author{Robert Graham}
\address{Fachbereich Physik, Universit\"at-GH, \\
         45117 Essen, Germany}

\maketitle
\begin{abstract}
       We investigate the hydrodynamic low-frequency oscillations of 
        highly degenerate Fermi gases trapped in anisotropic 
       harmonic potentials. Despite the lack of an obvious spatial symmetry 
       the wave-equation turns out to be separable in elliptical coordinates, 
       similar to a corresponding result established earlier for 
       Bose-condensates. This result is used to give the analytical solution of the anisotropic wave equation for the hydrodynamic modes.
\end{abstract}

\vskip2pc
\section{Introduction}\label{sec:1}

After the spectacular realization of Bose--Einstein condensation in evaporatively cooled magnetically trapped alkali vapors half a decade ago \cite{1} the experimental realization of a degenerate (i.e. mean particle distance smaller than the de Broglie wavelength) Fermi gases like $^{40}K$ \cite{2} and $^6Li$ \cite{3} has also been achieved. One of the most useful experimental tools to probe the properties of the Bose-condensates has been the excitation of their collective modes in the collisionless regime \cite{4,5}. It has turned out that within the dynamic Thomas-Fermi approximation these modes could be calculated analytically by solving the Gross-Pitaevskii equation, linearized around the static Thomas-Fermi solution. This was done first \cite{6,7} for isotropic parabolic traps, where the symmetries are sufficient for integrability, then for anisotropic axially symmetric parabolic traps \cite{6,8,9}, where an additional hidden symmetry could be identified \cite{8}, and finally also for completely anisotropic parabolic traps \cite{10}, where the resulting wave-equation could be related to various integrable dynamical systems and the Lam\'e equation.

In view of these successful experimental and theoretical developments for the Bose-systems, it seems to be of considerable interest to apply similar methods also to degenerate Fermi systems \cite{11,12}. On the one hand these systems seem to be very different from their bosonic counterparts. Indeed, barring the possibility of Cooper pairing at extremely low temperatures, condensates may not appear there. However, a degenerate Fermi gas may appear, instead. Therefore a great similarity between both kinds of systems remains, because in the strongly degenerate case the particles move in the trap-potential and strong mean-fields coming from their effective interactions, either with the condensate in the bosonic case, or via the Pauli exclusion-principle with the degenerate gas, in the case of fermions.

In the present paper our goal is to analyze the wave-equation of the collective modes of the degenerate Fermi gas in a completely anisotropic (i.e. with three
different trap-frequencies) parabolic trap. More symmetric special cases of this problem have already been analyzed in the literature.

For the special case of an isotropic parabolic trap the wave-equation for hydrodynamic modes in trapped Fermi gases was solved by Bruun and Clark \cite{12} both in the high- and low-temperature limit. In the high-temperature limit the effects of quantum statistics become irrelevant and one obtains the same results as for Bose gases, which were considered earlier, for isotropic traps, and even for some modes  in axially symmetric traps by Griffin et al. \cite{14}. Using a sum-rule approach Vichi and Stringari \cite{11} analyzed interaction effects on the collective oscillations of Fermi gases trapped in parabolic potentials of rotational or axial symmetry. Amoruso et al. \cite{13} also investigated the collective excitations of trapped degenerate Fermi-gases both in the hydrodynamic and the collisionless regime for isotropic and, for some dipolar and quadrupolar modes, also for axially symmetric parabolic traps. New in the present work is the treatment of the completely anisotropic case, where we follow the lines of our earlier treatment of the bosonic case \cite{10}. The paper is organized as follows.

In section \ref{sec:2} the hydrodynamic equations are formulated and the wave-equation for small density perturbations in the limit $T\rightarrow 0$ is derived.
In  section \ref{sec:3} the wave-equation is solved in the axially symmetric case by separation in cylindrical-elliptic coordinates. We also construct the conserved operator commuting with the wave-operator, whose presence is responsible for the integrability. In section \ref{sec:4} we proceed to the triaxially anisotropic case and separate the wave-operator in three-dimensional elliptic coordinates, exhibiting two conserved operators commuting with the wave-operator. Analytical expressions for the mode-frequencies and -eigenfunctions are given. Finally, we summarize our conclusions. 

\section{Equations of motion}\label{sec:2}
The hydrodynamic equations of motion of an ideal gas in local thermodynamic equilibrium with energy-density $\varepsilon (\vec{x},t)= \frac{3}{2} P (\vec{x}, t)$, where $P$ denotes the pressure, are given by \cite{15,14}

\begin{eqnarray}
\rho\frac{d\vec{u}}{dt} = -\frac{\vec{\nabla}P}{m}-\rho\frac{1}{m}
    \vec{\nabla} V (\vec{x})
\label{eq:(2.1)}
\end{eqnarray}
\begin{eqnarray}
\frac{dP}{dt}     =  -\frac{5}{3}\left(\vec{\nabla}\cdot\vec{u}P+\rho
     \vec{u}\cdot\vec{\nabla}V(\vec{x})\right)
\label{eq:(2.2)}
\end{eqnarray}
\begin{eqnarray}
\frac{d\rho}{dt}  =  -\rho\vec{\nabla}\cdot u
\label{eq:(2.3)}
\end{eqnarray}
with $d/dt = \partial/ \partial t + \vec{u}\cdot\vec{\nabla}$.

Here $\rho$ is the total density of the gas of fermions, which we assume to be present in $(2F+1)$ different hyperfine-states, which all see the same trapping potential and between which s-wave collisions are not suppressed by the Pauli-principle. We shall not consider here additional hydrodynamic modes connected with the internal degrees of freedom of the fermions, like spin-waves.

The trapping potential is assumed to have the form
\begin{eqnarray}
     V(\vec{x})=
    \frac{m}{2}\left( \omega_x^2 x^2 + \omega_y^2 y^2 + \omega_z^2 z^2\right).
\label{eq:(2.4)}
\end{eqnarray}
The number-density $\rho_0(\vec{x})$ in equilibrium is determined by the Fermi-Dirac distribution in the semi-classical approximation

\begin{eqnarray}
f_0(\vec{x},\vec{p})= \frac{1}{e^{\beta(\frac{p^2}{2m}+V(\vec{x})-\mu)}+1},
\label{eq:(2.5)}
\end{eqnarray}
where $\beta=1/(k_BT)$ and $\mu$ is the chemical potential. $f_0(\vec{x},\vec{p})$ reduces in the low-temperature limit to
\begin{eqnarray}
f_0(\vec{x},\vec{p})= \Theta 
     \left(\mu -\frac{p^2}{2m} - V(\vec{x})\right),
\label{eq:(2.6)}
\end{eqnarray}
where $\Theta(x)$ is the Heaviside jump-function.
Thus, summing over the hyperfine-states $\sigma$ we get
\begin{eqnarray}
\rho_0(\vec{x}) =\sum_\sigma \int\frac{d^3p}{(2\pi\hbar)^3}f_0 (\vec{x},\vec{p}),
\label{eq:(2.7)}
\end{eqnarray}
which , for $T\rightarrow 0$, reduces to

\begin{eqnarray}
\rho_0 (\vec{x}) = \frac{(2F+1)}{6\pi^2}
        \left[\frac{2m}{\hbar^2}
        \left(\mu - V (\vec{x})\right)\right]^{3/2}\Theta(\mu - V (\vec{x})),
\label{eq:(2.8)}
\end{eqnarray}
where $\Theta(.)$ is the Heaviside jump-function.
The degenerate Fermi-gas has a sharp boundary at $\mu=V(\vec{x})$ which forms a Fermi-ellipsoid for parabolic trapping potentials in complete analogy to the Thomas-Fermi ellipsoids formed by large condensates in such traps. Integrating over the whole cloud of particles we get
\begin{eqnarray}
N = \frac{(2F+1)}{6} \left( \frac{\mu}{\hbar\bar{\omega}}\right)^3
\label{eq:(2.9)}
\end{eqnarray}
with the geometric mean trap-frequency
\begin{eqnarray}
\bar{\omega} = (\omega_x \omega_y \omega_z)^{1/3}.
\label{eq:(2.10)}
\end{eqnarray}
The pressure $P_0(\vec{x})$ in equilibrium satisfies the hydrostatic condition

\begin{eqnarray}
\vec{\nabla} P_0(\vec{x}) = -\rho_0(\vec{x})\vec{\nabla} V (\vec{x}),
\label{eq:(2.11)}
\end{eqnarray}
from which
\begin{eqnarray}
P_0(\vec{x}) = \frac{2}{5}\left(\mu - V (\vec{x})\right)\rho_0(\vec{x})
\label{eq:(2.12)}
\end{eqnarray}
follows. The same result follows, for $T\rightarrow 0$, from
\begin{eqnarray}
   P_0(\vec{x})=
   \frac{2}{3}\epsilon_0 (\vec{x}) =
   \frac{2(2F+1)}{3} \int\frac{d^3k}{(2\pi)^3}
   \frac{\hbar^2k^2}{2m}
   f_0(\vec{x},\hbar \vec{k}).
\label{eq:(2.13)}
\end{eqnarray}
We can eliminate $\mu-V(\vec{x})$ from eqs.(\ref{eq:(2.12)}) and  (\ref{eq:(2.18)}) to obtain the equation of state
\begin{eqnarray}
   P_0 =\frac{1}{5} \frac{\hbar^2}{m}
        \left(\frac{6\pi^2}{2F+1}\right)^{2/3} \rho_0^{5/3}.
\label{eq:(2.14)}
\end{eqnarray}
Linearizing eqs.(\ref{eq:(2.1)}), (\ref{eq:(2.2)}), and (\ref{eq:(2.3)}) around equilibrium, with 
$\rho=\rho_0 +\delta\rho, P=P_0+\delta P_0$ we obtain
\begin{eqnarray}
    \rho_0(\vec{x})\frac{\partial\vec{u}}{\partial t}
  =  - \frac{1}{m}\vec{\nabla}\delta P
     - \delta\rho\frac{1}{m}\vec{\nabla} V(\vec{x}) 
\label{eq:(2.15)}
\end{eqnarray}
\begin{eqnarray}
  \frac{\partial\delta P}{\partial t} 
 = -\frac{5}{3}\vec{\nabla}\cdot\Bigl[P_0(\vec{x})\vec{u}\Bigr]
 - \frac{2}{3}\rho_0(\vec{x})\vec{u}\cdot\vec{\nabla} V (\vec{x}) 
\label{eq:(2.16)}
\end{eqnarray}
\begin{eqnarray}
  \frac{\partial\delta\rho}{\partial t}
=  - \vec{\nabla}\cdot\Bigl[\rho_0 (\vec{x})\vec{u}\Bigr].
\label{eq:(2.17)}
\end{eqnarray}
Eliminating $\delta P$ and $\delta\rho$ from these equations using the hydrostatic equilibrium condition relating $P_0(\vec{x})$ and $V(\vec{x})$ and the property $\rho_0(\vec{x}) = \rho_0(V(\vec{x}))$ we obtain \cite{14,12}
\begin{eqnarray}
   \frac{\partial^2\vec{u}}{\partial t^2}
  = \frac{5}{3m}\frac{P_0}{\rho_0}\vec{\nabla}\left(\vec{\nabla}\cdot
 \vec{u}\right)-\frac{2}{3m}\left(\vec{\nabla} V\right)\vec{\nabla}\cdot
 \vec{u} - \frac{1}{m}\vec{\nabla}\left(\vec{u}\cdot\vec{\nabla}V\right).
\label{eq:(2.18)}
\end{eqnarray}
Using in addition eq.(\ref{eq:(2.12)}) we obtain the wave-equation
\begin{eqnarray}
   \frac{\partial^2\vec{u}}{\partial t^2}
 = \frac{1}{m}\vec{\nabla} 
  \left[\frac{2}{3}(\mu-V(\vec{x}))\vec{\nabla}
  \cdot\vec{u} - \left(\vec{\nabla}\cdot V(\vec{x})\right)
  \cdot\vec{u}\right].
\label{eq:(2.19)}
\end{eqnarray}
Excluding time-independent 
flows, the gradient-form of the right-hand side permits the introduction of a velocity-potential
$\varphi$
\begin{eqnarray}
\vec{u} = -\vec{\nabla}\varphi
\label{eq:(2.20)}
\end{eqnarray}
in terms of which the wave-equation becomes
\begin{eqnarray}
   \frac{\partial^2\varphi}{\partial t^2} =
   \hat{G}\varphi = \frac{2}{3m}\left(\mu - V(\vec{x})\right)\nabla ^2
   \varphi - \frac{1}{m}\left( \vec{\nabla}V (\vec{x})\right) 
   \cdot\vec{\nabla}\varphi.
\label{eq:(2.21)}
\end{eqnarray}
From the continuity equation we can derive, with $\vec{u}=-\vec{\nabla}\varphi$ and $\partial\delta\rho/\partial t\ne 0$,
\begin{eqnarray}
   \delta \rho = \frac{3m}{2}
               \frac{\rho_0(\vec{x})}{\mu - V(\vec{x})}
               \frac{\partial\varphi}{\partial t}.
\label{eq:(2.22)}
\end{eqnarray}
Similarly, from the energy-balance equation it follows that
\begin{eqnarray}
   \delta P = m\rho_0(\vec{x})\frac{\partial\varphi}{\partial t},
\label{eq:(2.23)}
\end{eqnarray}
which implies the relation 
$\delta P=\frac{2}{3} (\mu-V(\vec{x}))\delta\rho$, and shows, together with eqs.(\ref{eq:(2.12)}) and (\ref{eq:(2.18)}), that
\begin{eqnarray}
   \delta P = 
   \frac{\partial P_0 (\rho_0)}{\partial\rho_0}\delta\rho.
\label{eq:(2.24)}
\end{eqnarray}
In the case of an isotropic trap the potential $V(\vec{x})$ becomes $V(\vec{x})=\frac{1}{2}m \omega_0^2 r^2$. Rescaling lengths by the Thomas-Fermi radius $r_{TF}=\sqrt{\frac{2\mu}{m\omega_0^2}}$ and separating the time-dependence by $\varphi(\vec{x},t)=e^{-i\omega t}\varphi(\vec{x})$ the wave-equation (\ref{eq:(2.21)}) in spherical coordinates 
$r,\phi, \theta$ simplifies to
\begin{eqnarray}
   -\omega^2 \varphi(r,\phi,\theta)=
   \omega_0^2\Bigl[
   \frac{1}{3}\frac{(1-r^2)}{r^2}\left(
   \frac{d}{dr}r^2\frac{d}{dr}-\nabla^2_{\theta,\phi}\right)
   - r\frac{d}{dr}\Bigr] \varphi(r,\phi,\theta).
\label{eq:(2.25)}
\end{eqnarray}
Here $\frac{1}{r^2}\nabla^2_{\theta,\phi}$ is the angular part of the Laplacian.
The separation ansatz $\varphi=r^l F_{n,l}(r^2)Y_{lm}(\theta\phi)$ with the spherical harmonics $Y_{lm}$, combined with the ansatz of a power-series of finite order for $F_{n,l}$
\begin{eqnarray}
      F_{nl}(r^2) = \sum_{k=0}^{n} C_k r^{2k},
\label{eq:(2.26)}
\end{eqnarray}
yields for the term of highest power $2k = 2n$ in $r$
\begin{eqnarray}
     \omega^2 r^{l+2n}
   = \omega_0^2\left[\frac{1}{3}\frac{d}{dr}r^2\frac{d}{dr}
   - \frac{l(l+1)}{3}+r\frac{d}{dr}\right] r^{l+2n},
\label{eq:(2.27)}
\end{eqnarray}
from which the mode-frequencies
\begin{eqnarray}
      \omega^2 = \frac{4}{3}\omega_0^2 \left(n^2+nl+2n+\frac{3l}{4}\right)
\label{eq:(2.28)}
\end{eqnarray}
follow, as first derived by Bruun and Clark \cite{12} and Amoruso et al. \cite{13}.
The polynomials $F_{nl}(x)$ are given by the hypergeometric functions \cite{12}
\begin{eqnarray}
      F_{nl}(x) = const \cdot \,_2F_1 (-n, l+2+n; l+\frac{3}{2}; x).
\label{eq:(2.29)}
\end{eqnarray}
In the following sections we consider the solution of eq.(\ref{eq:(2.21)})
 for axially symmetric and for triaxially anisotropic parabolic traps.

\section{Solution for axially symmetric traps}\label{sec:3}
In order to solve eq.(\ref{eq:(2.21)}) for axially symmetric trapping potentials
\begin{eqnarray}
      V(\vec{x})= \frac{1}{2}m\omega_{\bot}^2\left(x^2+y^2\right)
                + \frac{1}{2}m\omega_{\parallel}^2 z^2
\label{eq:(3.1)}
\end{eqnarray}
it is useful to introduce cylindrical-elliptic coordinates
\begin{eqnarray}
     \sqrt{x^2+y^2} = \sigma\sqrt{(1+\xi^2)(1-\eta^2)}, \quad
           z = \sigma\xi\eta,\quad
        \phi = arctan \frac{y}{x},
\label{eq:(3.2)}
\end{eqnarray}
 The focal length $\sigma$ of the Thomas-Fermi ellipsoid is
\begin{eqnarray}
     \sigma = \sqrt\frac{2\mu}{m}
          \sqrt{\frac{1}{\omega_{\bot}^2}- \frac{1}{\omega_{\parallel}^2}},
\label{eq:(3.3)}
\end{eqnarray}
its excentricity is $\epsilon = \sqrt{1-\omega_{\bot}^2/\omega_{\parallel}^2}$. The expressions as given apply for the case 
$\omega_{\bot}<\omega_z$. The case $\omega{\bot}\ge\omega_{\parallel}$
can be treated similarly with results which can also be obtained by the analytical continuation of the results given here, see \cite{8}.
The wave-equation can be transformed to these coordinates. Separating the time-dependence via
\begin{eqnarray}
   \varphi(\vec{x},t) = e^{-i\omega t}\varphi(\vec{x})
\label{eq:(3.3a)}
\end{eqnarray}
it takes the form
\[
   \frac{(1-\epsilon^2-\epsilon^2\xi^2)(1-\epsilon^2+\epsilon^2\eta^2)}
        {(1-\epsilon^2)(\xi^2+\eta^2)}
  \left\{\frac{2}{3}
     \left[
     \left(1+\xi^2\right)\frac{\partial^2}{\partial\xi^2}
     +
     \left(1-\eta^2\right)\frac{\partial^2}{\partial\eta^2}
     +
     2\xi\frac{\partial}{\partial\xi}-2\eta\frac{\partial}{\partial\eta}
     +
   \left( \frac{1}{1-\eta^2} -\frac{1}{1+\xi^2}\right)
          \frac{\partial^2}{\partial\phi^2}
     \right]\right.
\]
\begin{eqnarray}
    \left.
  - \frac{2\epsilon^2(1+\xi^2)}{1-\epsilon^2-\epsilon^2\xi^2}
         \xi
    \frac{\partial}{\partial\xi}
  + \frac{2\epsilon^2(1-\eta^2)}{1-\epsilon^2+\epsilon^2\eta^2}
         \eta
    \frac{\partial}{\partial\eta}
  +  \left( 
          \frac{1}{1-\epsilon^2-\epsilon^2\xi^2} 
       -  \frac{1}{1-\epsilon^2+\epsilon^2\eta^2}\right)
       \frac{2\omega^2}{\omega_{\parallel}^2}\right\} \varphi
  = 0
\label{eq:(3.4)}
\end{eqnarray}
In order to solve this equation we make the ansatz
\begin{eqnarray}
     \varphi = e^{im\phi} \varphi_{\xi}(\xi) \varphi_{\eta}(\eta),
\label{eq:(3.5)}
\end{eqnarray}
which permits the separation of variables and leads to the two equations
\begin{eqnarray}
   \left\{
   \frac{2}{3}
   \Bigl[
         (1+\xi^2)\frac{d^2}{d\xi^2}
       + 2\xi\frac{d}{d\xi}+\frac{m^2}{1+\xi^2}  \Bigr]
   -   \frac{2\epsilon^2(1+\xi^2)}{1-\epsilon^2-\epsilon^2\xi^2}
   \xi
       \frac{d}{d\xi} 
      + \frac{2\omega^2 / \omega_{\parallel}^2}{1-\epsilon^2-\epsilon^2\xi^2}
   \right\}
       \varphi_{\xi}(\xi) = B\varphi_{\xi}(\xi)
\label{eq:(3.6)}
\end{eqnarray}
\begin{eqnarray}
   \left\{
   \frac{2}{3}
   \Bigl[
         (1-\eta^2)\frac{d^2}{d\eta^2}
      - 2\eta \frac{d}{d\eta} - \frac{m^2}{1-\eta^2}  \Bigr]
     + \frac{2\epsilon^2(1-\eta^2)}{1-\epsilon^2+\epsilon^2\eta^2}
    \eta
       \frac{d}{d\eta}
      - \frac{2\omega^2 / \omega_{\parallel}^2}{1-\epsilon^2+\epsilon^2\eta^2}
   \right\}
       \varphi_{\eta}(\eta) = -B\varphi_{\eta}(\eta).
\label{eq:(3.7)}
\end{eqnarray}
Here $B$ is the separation constant. It is the eigenvalue of a conserved operator $\hat{B}$, commuting with the wave-operator $\hat{G}$,
\begin{eqnarray}
      [ \hat{B},\hat{G}  ] = 0.
\label{eq:(3.7a)}
\end{eqnarray}
It can be constructed from eqs.(\ref{eq:(3.6)}) and (\ref{eq:(3.7)}) by eliminating the eigenvalue $\omega^2$ of $\hat{G}$ from both equations and replacing the eigenvalue $m$ of the angular-momentum $\hat{L}$ around the $z$-axis by the operator $\hat{L_z} = -i \partial/\partial\phi$. We obtain in elliptic coordinates
\[
  \hat{B} = \frac{1}{\epsilon^2(\xi^2+\eta^2)}
  \left\{
    -\frac{2}{3}\left (1-\epsilon^2-\epsilon^2\xi^2\right)
     \frac{\partial}{\partial\xi} (1+\xi^2)\frac{\partial}{\partial\xi}
    + 2\epsilon^2(1+\xi^2)\xi\frac{\partial}{\partial\xi}    \right.
    \qquad\qquad\qquad\qquad\qquad
\]
\begin{eqnarray}
\qquad
  \left.
    - \frac{2}{3}(1-\epsilon^2+\epsilon^2\eta^2)\frac{\partial}{\partial\eta}
      (1-\eta^2)\frac{\partial}{\partial\eta}
    - 2\epsilon^2(1-\eta^2)\eta\frac{\partial}{\partial\eta}
    + \frac{2}{3}\left(\frac{1}{1+\xi^2}-\frac{1}{1-\eta^2}\right)
      \frac{\partial^2}{\partial\phi^2}
  \right\}
\label{eq:(3.8)}
\end{eqnarray}
Converted to the original coordinates $\hat{B}$ takes the simple form
\begin{eqnarray}
    \hat{B} = \frac{2}{3}
  \Bigl[ \left(\vec{x}\cdot\vec{\nabla}\right)
         \left(\vec{x}\cdot\vec{\nabla}+4\right)
       -  \frac{2\mu}{m\omega_{\bot}^2}
       \left(
       \frac{\partial^2}{\partial x^2}+\frac{\partial^2}{\partial y^2}
       \right)
       - \frac{2\mu}{m\omega_{\parallel}^2}
         \frac{\partial^2}{\partial z^2}
  \Bigr]
\label{eq:(3.9)}
\end{eqnarray}
The operator $\hat{B}$ can be freed from any dependence on the trap geometry \cite{Fliesser} (apart from the parabolic form of the trap-potential, which is essential for its existence) by rescaling the coordinates $\vec{x} \rightarrow\vec{\tilde{x}}$
\begin{eqnarray}
    \tilde{x} = x / \sqrt{\frac{2\mu}{m\omega_{\bot}^2}},\qquad
    \tilde{y} = y / \sqrt{\frac{2\mu}{m\omega_{\bot}^2}},\qquad
    \tilde{z} = z / \sqrt{\frac{2\mu}{m\omega_{\parallel}^2}}
\label{eq:(3.10)}
\end{eqnarray}
whence it becomes
\begin{eqnarray}
    \hat{B} = \frac{2}{3}
         \left[ 
         \left(  \vec{\tilde{x}}\cdot\vec{\tilde{\nabla}} \right)
         \left(  \vec{\tilde{x}}\cdot\vec{\tilde{\nabla}}+4\right)
       -  \tilde{\nabla}^2
         \right].
\label{eq:(3.11)}
\end{eqnarray}
which is rotationally invariant in the rescaled coordinates.
Apart from a trivial factor $2/3$ the operator $\hat{B}$ differs from its bosonic counterpart found in \cite{8} merely by a term proportional to $\vec{x}\cdot\vec{\nabla}$. 

The eigenvalue-problem of $\hat{B}$ can be easily solved simultaneously with that of $\hat{L}_z$ by the ansatz, in cylindrical coordinates $\rho, \phi, z$,
\begin{eqnarray}
    \varphi(\vec{x}) =
    e^{im\phi}\rho^{|m|}z^{\alpha} F_n^{(m,\alpha)}
    \left(  \rho^2, z^2\right)
\label{eq:(3.12)}
\end{eqnarray}
where $\alpha=0, 1$ determines the parity in $z$ and $F_n^{(m,\alpha)}$ is a polynomial of finite order $n$
\begin{eqnarray}
   F_n^{(m,\alpha)} =
       \sum_{k=0}^{n} \sum_{j=0}^{k} C_{k,j}^{(m,\alpha)}
                  \rho^{2j} z^{2(k-j)}.
\label{eq:(3.13)}
\end{eqnarray}

The condition that $n$ remains finite determines $B$ as
\begin{eqnarray}
  B = \frac{2}{3}
       \left( 2n+\alpha +|m|\right)
       \left( 2n+\alpha+|m|+4\right).
\label{eq:(3.14)}
\end{eqnarray}
Since $\hat{B}$ and $L_z$ both commute with the  wave-operator $\hat{G}$ they also have simultaneous eigenfunctions. Moreover, as we shall now show, the polynomial eigenfunctions of $\hat{B}$ with non-vanishing eigenvalue $B\not=0$ satisfy the requirement of particle-number conservation
\begin{eqnarray}
    \int d^3x\delta\rho(\vec{x},t) = 0.
\label{eq:(3.15)}
\end{eqnarray}
By eqs.(\ref{eq:(2.22)}), (\ref{eq:(2.8)}) and (\ref{eq:(3.3a)}), eq.(\ref{eq:(3.15)}) for 
$\omega\not= 0$ is equivalent to
\begin{eqnarray}
     -i\omega\int d^3x 
        \left(1-\frac{x^2}{a^2}-\frac{y^2}{a^2}-\frac{z^2}{c^2}\right)^{1/2}
        \varphi(\vec{x}) = 0  
\label{eq:(3.16)}
\end{eqnarray}
or, in rescaled space-coordinates
and for $\omega\ne 0$,
\begin{eqnarray}
     \int d^3\tilde{x}\left(
          1-\tilde{x}^2\right)^{1/2}\varphi(\vec{\tilde{x}}) = 0.
\label{eq:(3.17)}
\end{eqnarray}
Multiplying with $B\ne 0$ and inserting the operator $\hat{B}$ to replace its eigenvalue we can evaluate the left-hand side of eq.(\ref{eq:(3.17)}) by partial integration
\begin{eqnarray}
     B\int d^3\tilde{x}(1-\tilde{x}^2)^{1/2}\varphi
  &&=  \int d^3\tilde{x}(1-\tilde{x}^2)^{1/2}
  \left[
      \vec{\tilde{\nabla}} \cdot\vec{\tilde{x}}\vec{\tilde{x}} 
                           \cdot\vec{\tilde{\nabla}}
      - \vec{\tilde{\nabla}}^2 + \vec{\tilde{x}}\cdot\vec{\tilde{\nabla}}
  \right]\varphi\nonumber\\
  &&= - \int d^3\tilde{x}\left( \vec{\tilde{\nabla}}(1-\vec{x}^2)^{1/2}\right)
   \cdot \left( \vec{\tilde{x}}\vec{\tilde{x}}\cdot\vec{\tilde{\nabla}}
    - \vec{\tilde{\nabla}}\right) \varphi 
    + \int d^3\tilde{x}\left(1-\tilde{x}^2\right)^{1/2}\vec{\tilde{x}}
   \cdot \vec{\tilde{\nabla}}\varphi\nonumber\\
  &&= \int d^3\tilde{x}
     \frac{\tilde{x}^2-1}{\sqrt{1-\tilde{x}^2}}
   \vec{\tilde{x}}\cdot \vec{\tilde{\nabla}}\varphi 
   + \int d^3\tilde{x}(1-\tilde{x}^2)^{1/2}
      \vec{\tilde{x}}\cdot\vec{\tilde{\nabla}}\varphi\nonumber\\&&=0.
\label{eq:(3.18)}
\end{eqnarray}
We conclude that only the eigenfunctions with $\alpha=0=|m|=n$ must be excluded, because they correspond to $B=0$ and $\omega=0$. 

We must also guarantee momentum conservation of the modes, which means in a harmonic oscillator potential that the total momentum 
$\vec{P}(t)=\int d^3 x \rho_0(\vec{x})\vec{u}(\vec{x},t)$ must satisfy 
$\ddot{P}_i + \omega_i^2 P_i = 0$, where we denoted the three main frequencies $\omega_{x}$, $\omega_{y}$, $\omega_{z}$ by $\omega_i$, $i=1,2,3$ and the Cartesian components of $\vec{P}$ by 
$P_i$. Using the wave-equation (\ref{eq:(2.19)}) and our results for 
$\rho_0(\vec{x}), P_0(\vec{x})$ we can calculate
\begin{eqnarray}
      \left(
      \frac{\partial^2}{\partial t^2}+\omega_i^2  \right) 
      \int d^3 x\rho_0\vec{u}  =  +\frac{5}{3} \int d^3 x \nabla_i
      (
      P_0 \vec{\nabla}\cdot\vec{u})+\int d^3x\left[\vec{\nabla}\cdot(\rho_0\vec{u}\nabla_iV)-\nabla_i(\rho_0\vec{u}\cdot
\vec{\nabla}V\right],
\label{eq:(3.18a)}
\end{eqnarray}
which can give only surface-terms at the boundary of the degenerate Fermi-gas. These vanish for polynomial modes $\vec{u} = - \vec{\nabla}\varphi$ because 
$P_0 \sim(\mu - V(\vec{x}))^{5/2}$ and 
$\rho_0 \sim(\mu - V (\vec{x}))^{1/2}$ vanish at the boundary 
$\mu = V (\vec{x})$. Hence
\begin{eqnarray}
     \left(
     \frac{\partial^2}{\partial t^2}+\omega_i^2  \right)
     \int d^3 x\rho_0\vec{u}  =  0.
\label{eq:(3.18b)}
\end{eqnarray}

The wave-equation (\ref{eq:(2.21)}) leads to a recursion relation for the coefficients $C_{kj}^{(m,\alpha)}$ in a straight-forward manner. The coefficients $C_{nj}^{(m,\alpha)}$ corresponding to the highest total power $k=n$ satisfy a tridiagonal homogeneous set of equations
\begin{eqnarray}
     \sum_{j^{\prime}=0}^{n} M_{jj^{\prime}} C_{nj^{\prime}}^{(m,\alpha)}
     =
     \frac{3\omega^2}{2} \frac{m}{\mu} C_{nj}^{(m,\alpha)}
\label{eq:(3.19)}
\end{eqnarray}
with the matrix
\[
  M_{jj^{\prime}}  =  \delta_{j,j^{\prime}}
  \left[
  \frac{4j(j+|m|)+6j+3|m|}{a^2}  +
  \frac{4(n-j+\frac{\alpha}{2})(n-j+\frac{\alpha}{2}+1)}{c^2}
  \right]
\]
\begin{eqnarray}
  + \delta_{j,j^{\prime}+1}\cdot \frac{4}{a^2}
           \left( n-j+\frac{\alpha}{2}+1\right)
           \left(n-j+\frac{\alpha}{2}+\frac{1}{2}\right)
\label{eq:(3.20)}
\end{eqnarray}
\[
   + \delta_{j,j^{\prime}-1}\cdot \frac{4}{c^2}
     (j+1)(j+1+|m|)\qquad\qquad\qquad\qquad\qquad
\]
where we denote the radial and axial half-axis of the Fermi-ellipsoid as
$a=\sqrt{2\mu/m\omega_{\bot}^2}$ and $c=\sqrt{2\mu/m\omega_{\parallel}^2}$,
respectively. For any given pair of integers $n,|m|$ and parity $\alpha=0, 1$ this finite homogeneous set of $(n+1)$ equations is easily solved, in general numerically, and determines $(n+1)$ eigenfrequencies $\omega_j^{(n,|m|,\alpha)}$ labeled by the additional integer quantum number $j$ with $0\le j\le n$.

Simple solutions are obtained for $n=0$. We have already shown that $\alpha=m=0$ must be excluded if $n=0$, due to particle-number conservation. Therefore, the smallest frequencies are
\begin{eqnarray}
     \omega_0^{(0, 0, 1)}  =  \omega_{\parallel}\quad,\quad
     \omega_0^{(0, 1, 0)}  =  \omega_{\bot},
\label{eq:(3.21)}
\end{eqnarray}
which are the frequencies of the Kohn-modes \cite{16}. More generally
\begin{eqnarray}
     \omega_0^{(0,|m|,0)}  =  \sqrt{|m|}\,\,\omega_{\bot}\quad,\quad
     \omega_0^{(0,|m|,1)}  =  
                      \sqrt{\omega_{\parallel}^2+|m|\omega_{\bot}^2}
\label{eq:(3.22)}
\end{eqnarray}
with the (unnormalized) number-density modes
\begin{equation}
\delta\rho(\vec{x})= const\sqrt{1-\frac{\rho^2}{a^2}-\frac{z^2}{c^2}}e^{im\phi}\rho^{|m|}z^\alpha.
\end{equation}
The eigenfrequencies for $n=1$ are also easily obtained and given by
\begin{eqnarray}
     \omega_{1\choose 0}^{(1,|m|,0)} &&=  \sqrt{\frac{1}{3}}\left(5\omega_\bot^2(|m|+1)+4\omega_{\parallel}^2\pm\sqrt{(2|m|+5)^2\omega_\bot^4
+8\omega_{\parallel}^2(2\omega_{\parallel}^2-(|m|+4)\omega_\bot^2)}\right)^{1/2}\nonumber\\
     \omega_{1\choose 0}^{(1,|m|,1)}  &&=  
 \sqrt{\frac{1}{3}}\left(5\omega_\bot^2(|m|+1)+9\omega_{\parallel}^2\pm\sqrt{(2|m|+5)^2\omega_\bot^4
+36\omega_{\parallel}^2(\omega_{\parallel}^2-\omega_\bot^2)}\right)^{1/2}.\label{eq:(3.23)}
\end{eqnarray}
 The number-density modes take the form
\begin{equation}
\delta\rho^{(1,|m|,\alpha)}(\vec{x})= \sqrt{1-\frac{\rho^2}{a^2}-\frac{z^2}{c^2}}e^{im\phi}\rho^{|m|}z^\alpha
\left(C_{11}^{(m,\alpha)}\rho^2+C_{10}^{(m,\alpha)}z^2
+C_{00}^{(m,\alpha)}\right).
\end{equation}
The relation between the coefficients $C_{11}^{(m,\alpha)}$ and $C_{10}^{(m,\alpha)}$ for the $n=1$ modes can be obtained from eq.(\ref{eq:(3.19)}), 
\begin{equation}
C_{10}^{(m,\alpha)}=C_{11}^{(m,\alpha)}\frac{3\omega^2-(7|m|+10)\omega^2_\bot-\omega^2_\parallel\alpha}{2\omega^2_\bot(1+2\alpha)}
\end{equation}
and  
$C_{00}^{(m,\alpha)}$ is then most easily obtained from the recursion-relation
of the eigen-value equation for ${\hat B}$, which yields
\begin{equation}
C_{00}^{(m,\alpha)}=-\frac{3}{2}\frac{a^2(|m|+1)C_{11}^{(m,\alpha)}+c^2(1/2+\alpha)C_{10}^{(m,\alpha)}}{3+|m|+\alpha}.
\end{equation}
The results for the eigenfrequencies (\ref{eq:(3.23)}) for the special case $|m|=0$, 
$\alpha=0$ have already been obtained in \cite{13}.

\section{Triaxially anisotropic case}\label{sec:4}
Let us now turn to the case of a parabolic potential without any further symmetries. We shall analyze this case using the procedure applied to Bose-condensates in our previous paper \cite{10}.
Ellipsoidal coordinates $\lambda,\mu,\nu$ are introduced as functions of $x, y, z$ as the three roots of \cite{17}
with respect to $\rho$
\begin{eqnarray}
  \frac{x^2}{a^2+\rho}  +  \frac{y^2}{b^2+\rho} + \frac{z^2}{c^2+\rho}=1
\label{eq:(4.1)}
\end{eqnarray}
where $a\ge b\ge c$ are the semi-axis' of the Fermi-ellipsoid. We order the roots according to
\begin{eqnarray}
 -a^2\le\nu\le - b^2\le\mu\le -c^2\le\lambda\le 0.
\label{eq:(4.2)}
\end{eqnarray}
In order to avoid confusion the chemical potential is denoted as $\tilde{\mu}$ from now on. It is useful to define 
\begin{eqnarray}
  F(\rho)=
  \left( a^2+\rho\right)  \left( b^2+\rho\right) \left( c^2+\rho\right)
\label{eq:(4.3)}
\end{eqnarray}
The wave-equation can then be written as
\begin{eqnarray}
\frac{m}{{\tilde \mu}} a^2 b^2 c^2 \omega^2 \psi=
     -\frac{4\lambda\mu\nu}{(\lambda-\mu)(\mu-\nu)(\nu-\lambda)}
 \left\{ (\mu-\nu)  
    \left[   \frac{2}{3}F(\lambda)\frac{\partial^2}{\partial\lambda^2}
  +  \left(   \frac{F(\lambda)}{\lambda}+\frac{1}{3}F^{\prime}(\lambda)   \right)
              \frac{\partial}{\partial\lambda}
  \right] +  cyclic.    
 \right\}
\label{eq:(4.4)}
\end{eqnarray}
It can be separated like in the bosonic case \cite{10} with the result
\begin{eqnarray}
\left\{
       \frac{2}{3} 
       \lambda F(\lambda)\frac{\partial^2}{\partial\lambda^2}  +
       \left( F(\lambda)+\frac{1}{3}\lambda F^{\prime}(\lambda)\right)
             \frac{\partial}{\partial\lambda}
   \right\}   \varphi
    = \left\{
    \frac{m\omega^2}{\tilde{\mu}}  \frac{a^2b^2c^2}{4}
    +  \frac{A\lambda}{4}  +  \frac{B\lambda^2}{4}
       \right\}  \varphi  
\label{eq:(4.5)}
\end{eqnarray}
and two cyclic equations in $\mu,\nu$. $A$ and $B$ are separation constants. They are the eigenvalues of operators $\hat{A}$ and $\hat{B}$ which commute with the wave-operator $\hat{G}$. The existence of these conserved operators is the reason for the solvability of the wave-equation despite the absence of any apparent symmetries of the anisotropic potential.
In order to construct $\hat{A}$ and $\hat{B}$ we replace in (\ref{eq:(4.5)}) $A$ and $B$ by their operators and solve these three equations for $\hat{A}\varphi$ and $\hat{B}\varphi$. Afterwards we transform back to Cartesian coordinates. 
We obtain in this way after some calculation
\begin{eqnarray}
   \hat{A} = \frac{2}{3}
   \left\{ 
      \left[
         (b^2+c^2)(x^2-a^2)+a^2(y^2+z^2)    \right]
      \frac{\partial^2}{\partial x^2} + 2a^2y z 
      \frac{\partial^2}{\partial y\partial z}
    + 4 (b^2+c^2)x\frac{\partial}{\partial x}  + cyclic      \right\}     
\label{eq:(4.6)}
\end{eqnarray}
which differs from the bosonic case only by the trivial prefactor $2/3$ and the factor 4 instead of 3 in the last term. For the operator $\hat{B}$ we obtain
\begin{eqnarray}
   \hat{B} = \frac{2}{3}
   \left[
   \vec{x}\cdot\vec{\nabla}\left(\vec{x}\cdot\vec{\nabla}+4\right)
   - a^2 \frac{\partial^2}{\partial x^2} -b^2 \frac{\partial^2}{\partial y^2}
   - c^2 \frac{\partial^2}{\partial z^2}
   \right],
\label{eq:(4.7)}
\end{eqnarray}
which is a simple generalization of the result (\ref{eq:(3.9)}), and reduces again to (\ref{eq:(3.11)}) after  rescaling $x,y,z$ by $a,b,c$, respectively. The eigenfunctions of $\hat{B}$ and its eigenvalues are therefore still given by polynomials of finite total order $n_t$ in $x,y,z$ and
\begin{eqnarray}
      B = \frac{2}{3} n_t (n_t + 4), \qquad n_t=0, 1, 2,....
\label{eq:(4.8)}
\end{eqnarray}
The eigenfunctions of the wave-operator can now be obtained as in \cite{10} using the method for the solution of the Lam\'e  equation (see \cite{17}). The polynomial ansatz (with 
$n_t=2n+\alpha+\beta+\gamma;\,\,\alpha=0, 1;\,\,\beta=0, 1;\,\,\gamma=0, 1$)
\begin{eqnarray}
      \varphi = x^{\alpha}y^{\beta}z^{\gamma} \prod_{i=1}^n\varphi_i(\vec{x})
\label{eq:(4.9)}
\end{eqnarray}
with undetermined coefficients $\theta_i$, where
\begin{eqnarray}
      \varphi_i(\vec{x}) = 
      \left( 1-\frac{x^2}{a^2+\theta_i} 
              -\frac{y^2}{b^2+\theta_i}
              -\frac{z^2}{c^2+\theta_i}
      \right),
\label{eq:(4.10)}
\end{eqnarray}
inserted in the wave-equation $\hat{G}\varphi=-\omega^2\varphi$, yields after some rearrangement
\begin{eqnarray}
\frac{m}{{\tilde\mu}}\omega^2\varphi=&&\left[
      \frac{2\alpha}{a^2} + \frac{2\beta}{b^2}+ \frac{2\gamma}{c^2}
      - 4 \sum_{i=1}^{n} \frac{1}{\theta_i}
      \right]   \varphi\nonumber\\
&&+ \left(1-\frac{x^2}{a^2} -\frac{y^2}{b^2} -\frac{z^2}{c^2}\right)
            \sum_{i=1}^{n}
     \frac{\partial\varphi(\vec{x})}{\partial\varphi_i(\vec{x})}
     \left[
     \frac{4}{\theta_i} +\frac{4}{3}\frac{(2\alpha+1)}{a^2+\theta_i}
     +  \frac{4}{3}\frac{(2\beta+1)}{b^2+\theta_i}
     +  \frac{4}{3}\frac{(2\gamma+1)}{c^2+\theta_i}
     +  \frac{16}{3}\sum_{j=1\atop{(j\not=i)}}^{n} \frac{1}{\theta_i-\theta_j}
     \right].
\label{eq:(4.11)}
\end{eqnarray}
Here we define
\[
       \frac{\partial\varphi(\vec{x})}{\partial\varphi_i(\vec{x})}
    =  x^{\alpha}x^{\beta}x^{\gamma} \prod_{j=1\atop(j\not=i)}^{n}
       \varphi_j(\vec{x}).
\]
The ansatz for $\varphi$ is a solution if the second term on the left-hand side of eq.(\ref{eq:(4.11)}) vanishes. This is the case if, for $i=1,2, ...n$
\begin{eqnarray}
     \frac{6}{\theta_i} + \frac{4\alpha+2}{a^2+\theta_i}
   + \frac{4\beta+2}{b^2+\theta_i} + \frac{4\gamma+2}{c^2+\theta_i}
   + \sum_{j=1\atop(j\not=i)}^{n}  \frac{8}{\theta_i -\theta_j}
   = 0.
\label{eq:(4.12)}
\end{eqnarray}
It can be shown that all the roots of (\ref{eq:(4.12)}) are real \cite{10}.
The eigenfrequencies are then given by
\begin{eqnarray}
   \omega^2 = 
         \frac{2{\tilde\mu}}{m}
   \left(
         \frac{\alpha}{a^2} + \frac{\beta}{b^2} + \frac{\gamma}{c^2}
       - 2 \sum_{i=1}^{n} \frac{1}{\theta_i}
   \right).
\label{eq:(4.13)}
\end{eqnarray}
We consider the cases $n=0$ and $n=1$. For $n=0$ we find the 8 frequencies, for $\alpha,\beta,\gamma=0,1$
\begin{eqnarray}
    \omega =
      \sqrt{\frac{2{\tilde\mu}}{m}
           \left( \frac{\alpha}{a^2} + \frac{\beta}{b^2} 
                 + \frac{\gamma}{c^2}
           \right)}
\label{eq:(4.14)}
\end{eqnarray}
with
\begin{eqnarray}
    \varphi = x^{\alpha} x^{\beta} x^{\gamma}.
\label{eq:(4.15)}
\end{eqnarray}
The case $\alpha=\beta=\gamma=0$ with $\omega=0$ must be excluded however. The \underline{same} frequencies and modes are obtained in the bosonic case. For 
$n\ge 1$ the mode-frequencies in the bosonic and fermionic case differ, however. For $n=1$ eq.(\ref{eq:(4.12)}) reduces to a cubic equation for 
$\theta_1=\theta$
\begin{eqnarray}
   \left(    4 (\alpha+\beta+\gamma) + 10   \right)    
   (    \theta+a^2)    
   (    \theta+b^2)    
   (    \theta+c^2)    
           -
   \left[\left(    4\alpha+2\right)   a^2 
   \left(    \theta+b^2\right)    
   \left(    \theta+c^2\right)  + cyclic\right] = 0   
\label{eq:(4.16)}
\end{eqnarray}
and the mode-frequencies for the eight possible choices of the $\alpha=\beta=\gamma$ are
\begin{eqnarray}
  \omega =
      \sqrt{\frac{2{\tilde\mu}}{m}
           \left( \frac{\alpha}{a^2} + \frac{\beta}{b^2} 
                 + \frac{\gamma}{c^2} - \frac{2}{\theta}
           \right)}.
\label{eq:(4.17)}
\end{eqnarray}
In the general case $n>1$ the $\theta_i,\, i=1,2,...n$ can be determined as the equilibrium of the $n$-particle potential
\[
  V(\theta) =
          - \frac{3}{4}\sum_{i=1}^{n} ln|\theta_i|
          - \left(\frac{\alpha}{2}+\frac{1}{4}\right)
             \sum_{i=1}^{n} ln|a^2+\theta_i| 
             \qquad\qquad\qquad\qquad\qquad
\]
\begin{eqnarray}
    -  \left( \frac{\beta}{2}+\frac{1}{4}\right)
       \sum_{i=1}^{n} ln|b^2+\theta_i|
    -  \left(\frac{\gamma}{2}+\frac{1}{4}\right)
       \sum_{i=1}^{n} ln|c^2+\theta_i|
\label{eq:(4.18)}
\end{eqnarray}
\[
    -  \sum_{i=1}^{n} \,\,\sum_{j=i+1}^{n} ln|\theta_i -\theta_j|,
       \qquad\qquad\qquad\qquad\qquad\qquad\qquad\qquad
\]
which differs only in the prefactor of the first term on the right-hand side from the bosonic result (cf. eq. (4.24) of \cite{10}). To find the minima of the potential (\ref{eq:(4.18)}) is a problem in the 2-dimensional electrostatics (because of the logarithmic potential) of n+4 point-charges on a line: The equilibrium positions $\theta_i$ of the $n$-charges of unit strength between the 4 fixed charges of strength $3/4$ at $\theta=0, (\frac{\alpha}{2}+\frac{1}{4})$ at $\theta=-a^2, (\frac{\beta}{2}+\frac{1}{4})$ at $\theta=-b^2$, and $(\frac{\gamma}{2}+\frac{1}{4})$ at $\theta=-c^2$ determine the eigenfrequencies via eq.(\ref{eq:(4.13)}). The solutions can be classified by the numbers $n_1, n_2, n_3$ of charges between $-c^2$ and $0$, $-b^2$ and  $c^2$, and $-a^2$ and $-b^2$, respectively, which are the three quantum numbers labeling the eigenvalues and eigenfunctions. We shall not present numerical results here, since these depend on the ratios $a/b, a/c$ as an input.
For a further discussion we can refer to \cite{10}, however, where the bosonic case is analyzed further by numerically solving the equivalent electrostatic problem for a specific experimentally realized trap-geometry.

\section{Conclusions}\label{sec:5}
In the present paper we have analyzed the hydrodynamic equations of an ideal and highly degenerate Fermi-gas trapped in a generally anisotropic harmonic potential. Despite the absence of any obvious spatial symmetries in the triaxially anisotropic case, or of insufficient spatial symmetry in the axially symmetric case, the wave-equation turns out to be separable just like in the bosonic case. In fact, from the point of view of dynamical systems the characteristics of the wave-equation (i.e. the rays described by the Hamilton-Jacobi equation obtained in the eikonal limit \cite{19}) are the same for the bosonic and the fermionic case; i.e. the wave-equations for the two cases can be considered as different `quantizations' of the same `classical' dynamical system, which was analyzed in \cite{10}, and which turned out to be integrable.

The frequencies of the lowest-lying modes the $n=0$ modes (\ref{eq:(4.14)}), (\ref{eq:(4.15)}) (but not the mode-functions for $\delta\rho(\vec{x})$)
turn out to be the same in the bosonic and the fermionic case. The reason for this coincidence is the fact that the velocity-potential $\varphi$ of these modes satisfies $\nabla^2\varphi=0$, which implies the incompressibility condition $\vec{\nabla}\cdot\vec{u}=0$. For incompressional modes the only terms depending on the quantum-statistics drop out from the wave-equation (\ref{eq:(2.18)}) for the velocity-field, as was already noted by Griffin et al. \cite{14}. However, the frequencies of the higher-lying modes differ, in general, between the Bose- and the Fermi-systems. In the axially symmetric case we have given all the mode-frequencies with at most 1 node in radial or axial direction, and with arbitrary axial angular momentum. In the completely anisotropic case useful explicit results for compressional modes with $\vec{\nabla}\cdot\vec{u}\ne 0$ can only be given in numerical form.
However, as a central result of the present work we have reduced the numerical task to a very straight-forward one, namely to the 2-dimensional electrostatics of n movable point-charges of unit strength on a line in the field of
4 fixed point charges of given strength on the same line.
This establishes a strong analogy between the fermionic case and the bosonic case, which has been reduced to the same problem (with slightly different
strengths of the fixed charges) in \cite{10}. As discussed there
this mapping of the eigenvalue-problem permits a complete classification of the eigenvalues and the eigenfunctions by the three integer quantum-numbers,
which give the number of movable charges between each pair of nearest neighbors
of the 4 fixed charges.

The physical realization requires the simultaneous trapping of several hyperfine states, in order to allow for the interaction of the different fermionic species by collisions, which are necessary to maintain local thermodynamic equilibrium. Because the collision-rates in degenerate Fermi-gases are suppressed by a Fermi-blocking factor proportional to $(T/T_F)^2$ compared to the classical collision rates \cite{18} it is necessary to use atomic species with particularly large positive or negative s-wave scattering-lengths or to produce the latter by tuning the system close to a Feshbach resonance.

In view of the recent rapid experimental progress in this field \cite{2},\cite{3} it may be hoped that a measurement of these collective modes may be possible in the near future.

\acknowledgements

This work has been supported by a project of the Hungarian Academy
of Sciences and the Deutsche Forschungsgemeinschaft under Grant No. 436 UNG 113/144. A. Cs. 
would like to acknowledge support by the Hungarian National Scientific
Research Foundation under Grant No. AKP 98-20 2,2. R. G.  wishes to acknowledge support by the Deutsche
Forschungsgemeinschaft through the Sonderforschungbereich 237
"Unordnung and gro\ss e Fluktuationen".

\end{document}